\documentclass{article}
\usepackage{amssymb}
\usepackage{graphicx}
\begin{document}
\title{Muon Spin Rotation in Heavy Electron Pauli Limited Superconductors}

\author{Vincent P. Michal\\
Commissariat \`a l'Energie Atomique,\\
INAC/SPSMS, 38054 Grenoble, France}
\date{}
\maketitle

\begin{abstract}
The formalism for analyzing the magnetic field distribution in the vortex lattice of Pauli limited heavy electron superconductors is applied to the evaluation of the vortex lattice static linewidth relevant to Muon Spin Rotation ($\mu$SR) experiment.
On the basis of the Ginzburg-Landau expansion for the superconductor free energy we study the evolution with respect to external field of the static linewidth both in the limit of independent vortices (low magnetic field) and in the near $H_\mathrm{c2}^\mathrm{p}(T)$ regime by using an extension of the Abrikosov analysis to Pauli limited superconductors. We conclude that in the Ginzburg-Landau regime with Pauli limit, the electrodynamics of the vortex lattice predicts anomalous variations with applied field  of the static linewidth which is a result of the spin response contribution to screening supercurrents that dominates the usual charge response.
The model is proposed as a benchmark for comparison with possible other effects including vortex core localized states or interplay with magnetism.
\end{abstract}

\section{Introduction}

As an example of a superconductor in the Pauli limit, the heavy electron system CeCoIn$_5$ has special properties regarding its response to external magnetic field. Notably Muon Spin Rotation $\mu$SR experiment \cite{Spehling} has revealed anomalous variations of the vortex lattice static linewidth $\sigma_\mathrm{s}^\mathrm{VL}$ with respect to magnetic field oriented along the tetragonal crystal c-axis. In Ref. \cite{Spehling} the static linewidth was measured at temperature $T=20\mathrm{mK}$, showed an increase with applied field from zero field to about $95\%$ the upper critical field and eventually decreased just before the first order superconductor to metal transition. 
The decrease in $\sigma_\mathrm{s}^\mathrm{VL}$ with respect to external field usually observed and analysed \cite{Yaouanc} is the hallmark of a diminution in the vortex lattice local field contrast due to the decrease of the inter-vortex spacing with increasing field.

Here it is shown that in the Ginzburg-Landau regime an increasing behavior of the static linewidth is predicted. This results from the Zeeman interaction of the electron spin with the superconductor internal field, which dominates the usual charge response supercurrents.
As a result the field distribution is modified on a distance $\sim\xi$ ($\xi$ is the coherence length obtained in the Ginzburg-Landau formulation) from the center of each vortex \cite{Michal}. The existence of the effect was pointed out \cite{Houzet} in the context of magnetism of the FFLO (Fulde-Ferrel-Larkin-Ovchinnikov) state.
In parallel to this, a numerical approach to Eilenberger equations was undertaken \cite{Machida} and effects of strong Pauli paramagnetism were highlighted in the vortex lattice state of Pauli limited superconductors.

Let us first discuss qualitatively the properties of Pauli limited heavy electron superconductors \cite{Houzet, Michal}.
This class is characterized by a zero temperature Maki parameter here defined $\alpha_{M0}=H_\mathrm{c20}^\mathrm{orb}/H_\mathrm{c20}^\mathrm{p}>1$ (an alternative definition includes a $\sqrt{2}$ factor which is not assumed here for clarity). We set $H_\mathrm{c20}^\mathrm{orb}= \phi_0/\xi_0^2$ and $H_\mathrm{c20}^\mathrm{p}=T_c/\mu$, the zero temperature scales for orbital and Pauli limiting fields respectively (we use throughout units where $\hbar=c=1$). $\phi_0 =\pi/e\simeq 2.07\times10^{−7}\textrm{G.cm}^2$ is the vortex fluxoid quantum, $e$ the absolute value of the electron charge, $\xi_0=v_F/T_c$ the $T=0$ Cooper pair radius or coherence length, $v_F=k_F/m^\ast$ the Fermi velocity, $k_F$ the Fermi momentum, $m^\ast$ the renormalized electron mass, $T_c$ the superconductor critical temperature, $\mu=g\mu_B/2$ the electron magnetic moment absolute value, $g$ the Land\'e factor, $\mu_B=e/(2m)$ the Bohr magneton, and $m$ the electron bare mass.

There are three characteristic lengths in the problem: the zero temperature coherence length $\xi_0$ defined above, $L(H_\mathrm{c20}^\mathrm{p})=\sqrt{\mu \phi_0/T_c}$ the inter-vortex distance of a square vortex lattice in the Pauli limit at temperature $T=0$ and field $H_\mathrm{c20}^\mathrm{p}$ (more generally we note $L(B)=\sqrt{\phi_0/B}$ the inter-vortex spacing of a square vortex lattice with internal field $B$), and the London penetration depth $\lambda_L=\sqrt{m^\ast/(4\pi ne^2)}$ with $n$ the electron density in the superconductor (at T=0 and for a cylindrical fermi surface this is the electron density in the 2D metal $n=k_F^2/(2\pi l_c)$ with $l_c$ the spacing between the planes of the tetragonal crystal). Hence
\begin{equation}
 \alpha_{M0}=\left[\frac{L(H_{c20}^p)}{\xi_0}\right]^2=\frac{\mu\phi_0T_c}{v_F^2}\sim\frac{m^\ast T_c}{m E_F},
 \label{Maki}
\end{equation}
and we define the Ginzburg-Landau ratio
\begin{equation}
 \kappa=\frac{\lambda_L}{\xi_0}\sim\sqrt{\frac{m^\ast}{m(r_ek_F)}}\frac{T_c}{E_F},
 \label{kappa}
\end{equation}
where $r_e=e^2/m$ is the classical radius of the electron and $r_ek_F\sim 10^{-5}$. 

The orders of magnitudes are as follows. In a classical, non-heavy electron superconductor, $m^\ast\sim m$ and $E_F\sim 10^3 T_c$ give $\kappa\sim1$ and $\alpha_{M0}\sim10^{-3}$. In CeCoIn$_5$ however, $T_c\simeq2.3\mathrm{K}$, $\xi_0\sim 100\AA$, $\lambda_L\sim5000\AA$ yield $m^\ast\sim300 m$, $E_F\sim 100 T_c$, $\kappa\sim50$, and $\alpha_{M0}\sim3$, which is as we shall see the origin of special magnetic properties of the vortex lattice. The large Ginzburg-Landau parameter implies \cite{deGennes} at $T=0$ the ratio between the field at which the first vortex nucleates in the bulk of the sample and the orbital upper critical field $H_\mathrm{c10}/H_\mathrm{c20}^\mathrm{orb}\sim \ln(\kappa)/\kappa^2\ll 1$, hence $B\simeq H$ for a broad magnetic field range. 
In a Pauli limited superconductor, $H_\mathrm{c10}/H_\mathrm{c20}^\mathrm{p}\sim(r_ek_F)(E_F/T_c)\ln(\kappa)\sim 10^{-3}$ and the same property applies.

We now study the electrodynamics of the vortex lattice which results from such large values for parameters (\ref{Maki}) and (\ref{kappa}). The vortex lattice static linewidth is defined as
\begin{equation}
\sigma_\mathrm{s}^\mathrm{VL}=\frac{\gamma_\mu}{\sqrt{2}}\sqrt{\overline{\delta h(\mathbf{r})^2}},
\label{sigma}
\end{equation}
where $\gamma_\mu=2\pi\times 135.5342\textnormal{MHz/T}$ is the muon gyromagnetic ratio, $h(\mathbf{r})$ is the component of the internal local field parallel to the applied field $H$, $\delta h(\mathbf{r})=h(\mathbf{r})-B$, the macroscopic internal field (or induction) $B=\overline{h(\mathbf{r})}$, and overline means averaging over a vortex lattice unit cell.    
Eq. (\ref{sigma}) can be expressed as a sum involving all order Fourier components $F_{mn}$ of the field distribution in the vortex lattice
\begin{equation}
\sigma_\mathrm{s}^\mathrm{VL}=\frac{\gamma_\mu}{\sqrt{2}}\sqrt{\sum_{(m,n)\neq(0,0)}(F_{mn})^2}.
\label{sigma2}
\end{equation}
The components $F_{mn}$ are called vortex lattice form factors \cite{Michal,FormFactors} in the context of Small Angle Neutron Scattering (SANS) experiment \cite{Note}.

\section{Muon static linewidth in the low-field-high-temperature regime}

\begin{figure}[t]
\centering
\includegraphics[width=12cm]{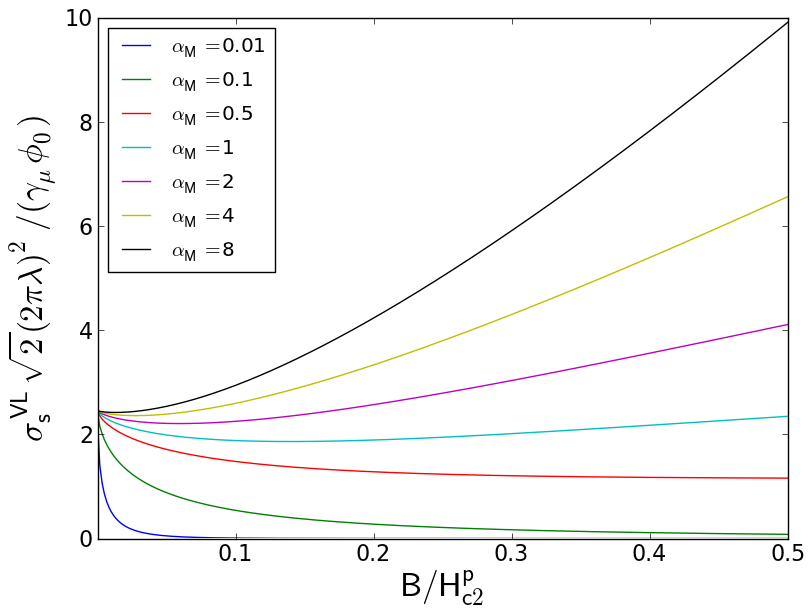}
\caption{Variations of the dimensionless $\mu$SR static linewidth  $\sigma_\mathrm{s}^\mathrm{VL}\sqrt{2}(2\pi\lambda)^2/(\gamma_\mu\phi_0)$ where $\sigma_\mathrm{s}^\mathrm{VL}$ is taken from Eq. (\ref{sigma2}) and the form factors from Eq. (\ref{f}). Different values for the temperature Maki parameter (\ref{MakiT}) where used as indicated in the legend.}
\label{Plot1}
\end{figure}

Here we use results of the Ginzburg-Landau formulation \cite{Houzet,Michal} to evaluate the static linewidth Eq. (\ref{sigma2}). The near-$T_c$ Ginzburg-Landau regime in the Pauli limit is accessible since the crossover temperature $T^\ast$ from orbitally limited to Pauli limited superconductivity is in the range $(T_c-T^\ast)/T_c\sim1/\alpha_{M0}^2$ (we shall see below this follows from the relation $H_\mathrm{c2}^\mathrm{orb}(T)/H_\mathrm{c2}^\mathrm{p}(T)\sim\alpha_{M0}\sqrt{1-T/T_c}$). 
In the independent vortex approximation (low magnetic field) and high-$\kappa$ limit the form factors can be decomposed as a sum of two distinct contributions \cite{Michal}
\begin{equation}
F_{mn}=F_{mn}^\mathrm{orb} + F_{mn}^\mathrm{Z}.
\label{F}
\end{equation}
The first term is the usual charge response which gives rise to orbital supercurrents. This writes \cite{Michal}
\begin{equation}
 F_{mn}^\mathrm{orb} = \frac{B\xi_v}{q_{mn}\lambda^2}K_1 (q_{mn}\xi_v),
 \label{Forb}
\end{equation}
where $\xi_v = \sqrt{2}\xi$ is a variational parameter which minimizes the superconductor free energy, $q_{mn}=[2\pi/L(B)](m^2+n^2)^{1/2}$ in a square vortex lattice, and $K_n(z)$ is the n$^\mathrm{th}$ order modified Bessel function of the second kind (or MacDonald function) \cite{Abramowitz}.
The near-$T_c$ coherence length and penetration depth depend on the symmetry of the superconducting gap \cite{Michal}. The expressions for d-wave pairing are 
\begin{equation}
 \frac{1}{\xi^2}=\frac{32\pi^2T_c^2}{7\zeta(3)v_F^2}\Big[\frac{T_c-T}{T_c}-7\zeta(3)\Big(\frac{\mu B}{2\pi T_c}\Big)^2\Big],
\label{xi}
\end{equation}
and
\begin{equation}
 \frac{1}{\lambda^2}=\frac{16}{3}\pi e^2v_F^2N_0\Big[\frac{T_c-T}{T_c}-7\zeta(3)\Big(\frac{\mu B}{2\pi T_c}\Big)^2\Big],
\label{lambda}
\end{equation}
where $N_0$ is the density of states with dimension $[\textnormal{Energy}\times\textnormal{Volume}]^{-1}$, $\zeta(z)$ is the Riemann zeta function, $\zeta(3)\simeq1.20$. The field at which these two lengths diverge is defined as the near-$T_c$ Pauli limit upper critical field
\begin{equation}
 H_\mathrm{c2}^\mathrm{p}=2\pi \frac{T_c}{\mu}\sqrt{\frac{T_c-T}{7\zeta(3)T_c}}.
 \label{Hc2p}
\end{equation}
The near-$T_c$ Zeeman spin contribution in Eq. (\ref{F}) for superconducting gap with d-wave symmetry reads \cite{Michal}
\begin{equation}
 F_{mn}^\mathrm{Z}=\frac{28\zeta(3)v_F^2N_0}{3\phi_0}\Big(\frac{\mu B}{T_c}\Big)^2K_0(q_{mn}\xi_v).
 \label{FZ}
\end{equation}
We scale the internal field, form factors and coherence length such that (\ref{F}) in dimensionless units evaluates to
\begin{equation}
 f_{mn}=\frac{q}{m^2+n^2}K_1(q)+4\pi b^2 K_0(q).
 \label{f}
\end{equation}
Here $b=B/H_\mathrm{c2}^\mathrm{p}$, 
\begin{equation}
 q=q_{mn}\xi_v=\sqrt{\frac{\pi\sqrt{7\zeta(3)}}{2}\frac{b}{\alpha_{M}}(m^2+n^2)},
\end{equation}
with
\begin{equation}
 \alpha_M=\alpha_{M0}\sqrt{1-\frac{T}{T_c}},
  \label{MakiT}
\end{equation}
$\alpha_{M0}$ is given in Eq. (\ref{Maki}), and $f_{mn}=F_{mn}(2\pi\lambda)^2/\phi_0$.
Near $T_c$ the dimensionless form factors (\ref{f}) take on a simple, universal form where only remains the parameter $\alpha_M$ controlling the relative contributions of the spin response with respect to the charge response.
The static linewidth variations in the independent vortex limit with dimensionless internal field $b$ and different values for $\alpha_M$ are shown in Fig. \ref{Plot1}. Observe the low field regime were all curves meet which follows from the limit
\begin{eqnarray}
 \nonumber\sqrt{\sum_{(m,n)\neq(0,0)}(f_{mn})^2}&\to& \sqrt{\sum_{(m,n)\neq(0,0)}\frac{1}{(m^2+n^2)^2}}\simeq2.455\\
&\textnormal{ as }b\to0.&
\end{eqnarray}

\begin{figure}[t]
\centering
\includegraphics[width=12cm]{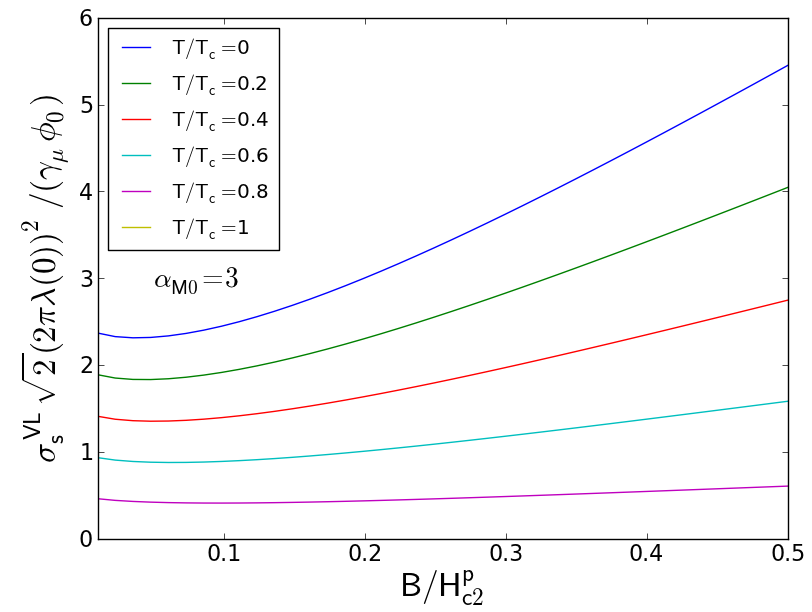}
\caption{Variations of the static linewidth $\sigma_\mathrm{s}^\mathrm{VL}\sqrt{2}[2\pi\lambda(0)]^2/(\gamma_\mu\phi_0)$ with field at different temperatures. We considered the parameter $\alpha_{M0}=3$.}
\label{Plot1-2}
\end{figure}

We now turn to the effect of temperature on the form factors Eq. (\ref{F}). We fix the value $\alpha_{M0}=3$ and plot $\sigma_\mathrm{s}^\mathrm{VL}\sqrt{2}[2\pi\lambda(0)]^2/(\gamma_\mu\phi_0)$ for different $T/T_c$, where $\lambda(0)$ is the Ginzburg-Landau penetration depth Eq. (\ref{lambda}) taken at $T=0$. The results are shown in Fig. \ref{Plot1-2}. Notice we have extended the temperature domain to very low $T/T_c$, which is not formally justified as it is in the near-$T_c$ region but is expected to give qualitatively meaningful variations. 

The MacDonald functions assume the limits $K_0(q)\to-\ln(q/2)-C$ and $K_1(q)\to1/q$ as $q\to0$ where $C\simeq0.5772$ is the Euler constant. Having SANS experiment in mind in the large-$\alpha_M$ limit it is useful considering Eq. (\ref{f}) with $(m,n)=(1,0)$
\begin{equation}
 f_{10}=1-2\pi b^2\ln\left(\frac{\pi\sqrt{7\zeta(3)}}{8\alpha_M}be^{2C}\right).
\end{equation}

\section{Muon static linewidth close to the second-order transition critical field}

The form factors as given in Eqs. (\ref{Forb}) and (\ref{FZ}) are found in the independent vortex approximation. The derivation does not work near the transition to the non-superconducting metal. In the high-field limit close to the transition line, the main source of magnetic field inhomogeneity in the vortex lattice comes from the Zeeman spin response \cite{Houzet,Michal} 
\begin{equation}
 \delta h(\mathbf{r})=-4\pi\varepsilon\left(|\Delta(x,y)|^2-\overline{|\Delta(x,y)|^2}\right),
\end{equation}
where
\begin{equation}
 \varepsilon=\frac{N_0\mu}{2\pi T}\Im\mathfrak{m}\Psi^{(1)}\Big(\frac{1}{2}-i\frac{\mu B}{2\pi T}\Big),
\end{equation}
by overlining we again mean averaging over a vortex lattice unit cell, and $\Psi^{(n)}(z)$ is the polygamma function \cite{Abramowitz} of order $n$.  
In a square vortex lattice the Fourier decomposition of the square of the gap magnitude reads \cite{Michal}
\begin{eqnarray}
 \nonumber|\Delta(x,y)|^2&=&\overline{|\Delta(x,y)|^2}\sum_{m,n=-\infty}^{+\infty}(-1)^{m+n+mn}\\&&\times e^{-\frac{\pi}{2}(m^2+n^2)}e^{2\pi imx/a}e^{2\pi iny/a}.
\end{eqnarray}
Therefore the form factors corresponding to Bragg peaks with indices $(m,n)\neq(0,0)$ take on the form
\begin{equation}
 F_{mn}=-4\pi\varepsilon\overline{|\Delta(x,y)|^2}(-1)^{m+n+mn}e^{-\frac{\pi}{2}(m^2+n^2)},\label{FF}
\end{equation} 
and the vortex lattice static linewidth simply reads
\begin{equation}
 \sigma_\mathrm{s}^\mathrm{VL}=\frac{4\pi s}{\sqrt{2}}\gamma_\mu\varepsilon\overline{|\Delta(x,y)|^2},
 \label{sigmasA}
\end{equation}
where
\begin{equation}
 s=\sqrt{\Big(\sum_{n=-\infty}^{+\infty}e^{-\pi n^2}\Big)^2-1}\simeq0.4247.
\end{equation}

\begin{figure}[t]
\centering
\includegraphics[width=12cm]{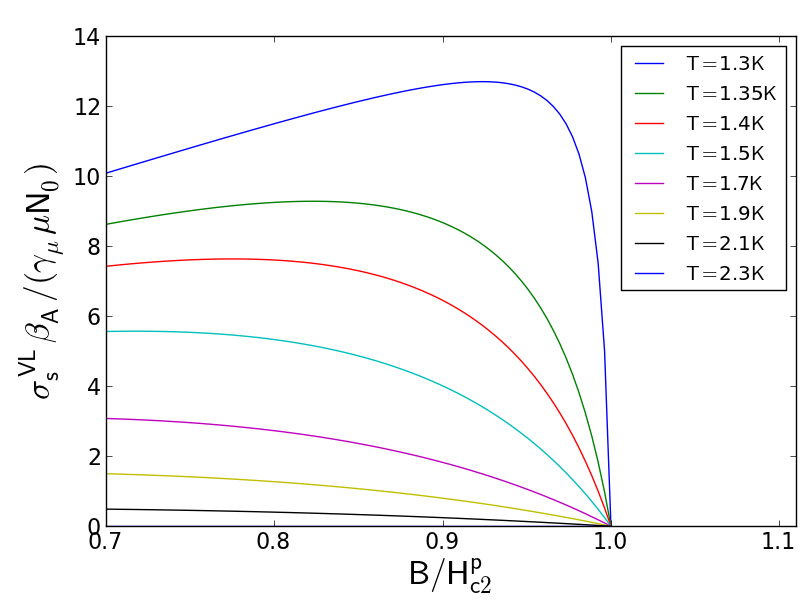}
\caption{$\mu$SR static linewidth close to the second-order transition line $H_\mathrm{c2}^\mathrm{p}(T)$ as obtained from Abrikosov's analysis in the Pauli limit Eq. (\ref{sigmasA}) for a temperature range as indicated in the legend. We have scaled the internal field with respect to $H_\mathrm{c2}^\mathrm{p}(T)$ here defined as the curve solution of $\alpha(T,B)=0$ \cite{Houzet,Michal}. Notice the rapid increase in the absolute value of the slope of $\sigma_\mathrm{s}^\mathrm{VL}(B)$ while approaching the first order transition at $T/T_c\simeq0.56$ and $\mu H_\mathrm{c2}^\mathrm{p}/T_c\simeq1.07$.}
\label{Plot2}
\end{figure}
Eq. (\ref{sigmasA}) shows explicitly that the vortex lattice contribution to the static linewidth vanishes when the transition is of the second order but shows a discontinuity where the transition is of the first order. In the former case, the gap average is \cite{Houzet,Michal}
\begin{equation}
\overline{|\Delta(x,y)|^2}=\frac{|\alpha|}{2\beta_A\beta}.
\end{equation}
Here 
\begin{equation}
\alpha=N_0\left[\ln\left(\frac{T}{T_c}\right)+\Re\mathfrak{e}\Psi\left(\frac{1}{2}-i\frac{\mu B}{2\pi T}\right)-\Psi\left(\frac{1}{2}\right)\right],
\end{equation}
and
\begin{equation}
 \beta=-\frac{3N_0}{64\pi^2T^2}\Re\mathfrak{e}\Psi^{(2)}\left(\frac{1}{2}-i\frac{\mu B}{2\pi T}\right)
\end{equation}
are the quadratic and quartic coefficients of the Ginzburg-Landau free energy \cite{Houzet, Michal} respectively, and $\Psi(z)$ is the digamma function \cite{Abramowitz}. $\beta_A=\overline{|\Delta(x,y)|^4}/\overline{|\Delta(x,y)|^2}^2$ is the Abrikosov parameter, it is $\beta_A^\square=1.18$ for a square vortex lattice and $\beta_A^\triangle=1.16$ for a triangular lattice. Then it follows
\begin{equation}
 \sigma_\mathrm{s}^\mathrm{VL}=\frac{2\pi s\gamma_\mu}{\sqrt{2}\beta_A}\frac{|\alpha|\varepsilon}{\beta},
\end{equation}
which is shown in Fig. \ref{Plot2}.

\section{Conclusion}

On the basis of the Ginzburg-Landau expansion for the superconductor free energy in the Pauli limit, we have studied the evolution with respect to external field of the muon spin rotation vortex lattice static linewidth both in the limit of independent vortices (low magnetic field) near $T_c$, and in the near $H_\mathrm{c2}^\mathrm{p}(T)$ regime.
In the first case, we have found a simple form for the total form factor which is a function of the internal field scaled with the temperature dependent upper critical field in the Pauli limit and includes a single parameter $\alpha_M=\alpha_{M0}\sqrt{1-T/T_c}$ with $\alpha_{M0}=\mu \phi_0T_c/v_F^2$. In the regime near $H_\mathrm{c2}^\mathrm{p}(T)$ we have applied an extension of the Abrikosov analysis to Pauli limited superconductivity and observed the approach from second-order to first-order transition to the metal (this occurs at $T/T_c\simeq0.56\mathrm{K}$ and $\mu H_\mathrm{c2}^\mathrm{p}/T_c\simeq1.07$) with a sudden raise in the absolute value of the slope of $\sigma_\mathrm{s}^\mathrm{VL}(B)$ while approaching $H_\mathrm{c2}^\mathrm{p}(T)$.
Such an analysis allows a simple modelling of the effect of heavy electron superconductor strong paramagnetism on the vortex lattice electrodynamics. 
It is proposed as a benchmark for studying new puzzling vortex lattice properties in CeCoIn$_5$ \cite{Michal2}.

\section{Acknowledgment}
I would like to thank V. P. Mineev for support and L. I. Glazman for stimulating discussion.


\begin{thebibliography}{99}
\bibitem{Note} The measurement of the first order form factor $F_{10}$ at $T = 50\mathrm{mK}$ \cite{FormFactors} has revealed a similar behavior to the one obtained in $\mu$SR experiment \cite{Spehling}: $F_{10}$ increases with field up to $4.7\,\textrm{T}$, and eventually decreases while approaching the (first order) superconductor to metal transition.
\end{thebibliography}
\end{document}